# Twist-bend nematics and heliconical cholesterics: a physico-chemical analysis of phase transitions and related specific properties

Natalia A. Kasian[1], Longin N. Lisetski[1], Igor A. Gvozdovskyy[2*]

[1]*Institute for Scintillation Materials, National Academy of Sciences of Ukraine, Kharkiv, Ukraine;* [2]*Institute of Physics, National Academy of Sciences of Ukraine, Kyiv, Ukraine;*

Institute of Physics, National Academy of Sciences of Ukraine, Prospekt Nauki 46, Kyiv, Ukraine; 03028, Ukraine, telephone number: +380 44 5250862, [*]E-mail: igvozd@gmail.com

# Twist-bend nematics and heliconical cholesterics: a physico-chemical analysis of phase transitions and related specific properties

Certain nematic liquid crystals, *e.g.*, those formed by banana-shaped molecules, can exhibit a low-temperature twist-bend nematic ($N_{tb}$) phase. Upon addition of chiral dopants, a chiral version of the twist-bend phase ($N^*_{tb}$) can be observed below the conventional chiral nematic ($N^*$) phase, while under electric field the $N^*$ phase is transformed into a specific state known as "heliconical cholesteric". In this work, we have studied the well-known $N_{tb}$-forming system CB7CB:CB6OCB with addition of 5CB and chiral dopants (ChD) using optical microscopy, differential scanning calorimetry (DSC), and measurements of temperature-dependent optical transmission using optical microscopy. Effects of 5CB and different chiral dopants (of steroid or non-steroid nature) upon thermal stability of $N_{tb}$ phase were analyzed, and selective reflection in the visible range could be observed in the $N^*$ phase with a sufficiently high ChD content. As a peculiar feature, sharp unwinding of the cholesteric helix was observed at lower temperatures when approaching the $N^*_{tb}$ phase, which appears to be similar to the helix unwinding close to the smectic-A transition.

## Introduction

Among various structural types of mesogenic molecules capable of forming various types of liquid crystal phases, an interesting example is presented by so-called banana-shaped molecules. These molecules are achiral, but under certain geometrical constraints they can give rise to chiral supramolecular arrangements. [1,2] At first, the conditions for observing such effects were limited to the translationally ordered smectic phases formed by banana-shaped molecules (the effects of chirality enhancement induced upon addition of such molecules to the already existing helically twisted phases [3,4] are of different nature and are most probably caused by specific intermolecular interactions). Recently, however, it was shown that some banana-shaped molecules consisting of two identical or near-identical mesogenic units connected by an odd-numbered alkyl or alkoxy chain can form, at temperatures below their nematic range, another distinct liquid crystalline phase – a phase with no smectic-like translational ordering, but with a rather peculiar kind of molecular arrangement with periodic twist and bend deformation of the director distribution. Such phase seems to be first reported in [5], and was later generally recognized as the "twist-bend nematic phase" ([6-10] and references therein) and designated as $N_{tb}$. In the $N_{tb}$ phase, the director forms a conical helix, and chirality appears spontaneously with equal fractions of helically-ordered "domains" of opposite handedness from an achiral liquid crystal at the molecular level. The "pitch" (*i.e.*, the period related to these helices) is typically around 10 nm, just a few molecular lengths, i.e., about 2-4 orders of magnitude shorter than pitches typical for cholesterics or chiral smectics $C^{*}$.

To move towards eventual practical applications of the twist-bend materials, one should obtain multi-component mixtures possessing the $N_{tb}$ phase in the desired temperature ranges, as it is routinely performed with conventional nematics and cholesterics. In this respect, the first basic steps were taken in [11], where temperatures and enthalpies of phase transitions were listed for a set of 1,7-bis(4-cyanobiphenyl-4′-yl) alkanes (CB*n*CB) and 1-(4-cyanobiphenyl-4′-yloxy)-ω-(4-cyanobiphenyl-4′-yl) alkanes (CB*m*OCB) with odd *n* and even *m* values; the latter condition ensured the bend-shaped molecular geometry. Basing on DSC thermograms, binary phase diagrams were constructed for several pairs of these compounds. The nematic to isotropic transition temperatures ($T_i$) varied almost linearly with concentration, while for the N–$N_{tb}$ transitions, a slight upward curvature was noted on the corresponding plots.

Multicomponent systems forming $N_{tb}$ phase were also studied in [12], where, apart from mixtures of two "twist-bend nematogens", the effects of introducing chiral dopants (ChD) on the N and $N_{tb}$ phases were reported. It appeared that $N_{tb}$ phase generally persisted up to 10% of ChD, with blue phase formation observed close to $T_i$. In [13], a general situation of twist-bend phase of chiral materials was considered, with mixtures of CB7CB and an intrinsically chiral bent-shaped mesogen showing a continuous range of $N_{tb}$ phase below the conventional $N^*$ phase. This phase (which should logically be denoted as $N^*_{tb}$) is optically positive (while $N^*$ phase is optically negative), and is characterized by a periodicity ("pitch") about 50 nm, which is by almost an order less that the values typical for standard $N^*$ phases.

Several recent papers [14-17] focused on the study of the $N^*$ phases in mixtures comprising bend-shaped molecules of CB$n$CB type and chiral dopants. Such systems under electric field undergo transition to a peculiar cholesteric state with oblique helicoidal structure denoted in many papers as $Ch_{OH}$. Application of the electric field to a cell filled with such material leads, above a certain threshold, to a well-defined picture of selective reflection bands, with the helical pitch $p$ (and the wavelength of visible colors) decreasing with increasing voltage; thus, broad possibilities appear for various electro-optical applications. The broadening of the temperature range can be achieved with addition of "standard" nematic components to the initial CB$n$CB-based mixture. [18] A moment of interest is that in all these papers there is no mentioning of the $N_{tb}$ phase – though it is absolutely clear that this is a natural low-temperature state of the mixtures studied. It is natural to assume that physico-chemical aspects of $Ch_{OH}$ mixtures in technological applications to be developed cannot be fully mastered without profound knowledge of the eventual low-temperature $N_{tb}$ state of such systems. (In a certain way, it resembles conventional cholesterics used as temperature sensors – the physical mechanism of helix unwinding cannot be understood without accounting for the low-temperature smectic-A phase forming in such systems).

An interesting observation is that reported $N^*_{tb} \rightarrow N^*$ transition temperatures were higher for the chiral dimers than seen for the racemic counterparts [19] (which would be quite logical, since spontaneous short-range chiral domains should be stabilized when combined into a joint macroscopic helix with a defined twist sense). This suggests that studies of mixtures including chiral dopants of different chemical nature could yield useful insight on the thermal stability of the $N_{tb}$ phases. Also, a detailed study of the effects of standard "rod-like" liquid crystal components (using the approach of [18]) could help in identifying ways to obtain broad range mixtures forming $Ch_{OH}$ and $N^*_{tb}$ phases.

Thus, in this manuscript our idea was to add various components to the known systems eventually forming $Ch_{OH}$ and $N_{tb}$ phases, to check (and, if possible, explain) the effects of these additives on phase transitions and properties of these phases by means of various methods such as DSC, polarizing optical microscopy, optical transmission spectra, *etc.*

**Experiment**

*2. Materials and methods*

To study the effects of the nematic 5CB and various ChDs on the properties of twist-bend/chiral twist-bend nematic ($N_{tb}/N^*_{tb}$) mixtures, we used a base $N_{tb}$ mixture [11,20] of 1,7-bis-4-(4-cyano-biphenyl)heptane (CB7CB) and 1-(4-cyanobiphenyl-4-yloxy)-6-(4-cyanobiphenyl-4-yl)hexane (CB6OCB) in the ratio of (1:1) obtained from Synthon

Chemicals GmbH & Co (Wolfen, Germany). It is known that CB7CB possesses a uniaxial nematic N phase in a temperature range of 103 - 116 °C, [5,11,21] with the elastic constants $K_1 \approx 8$ pN, $K_2 \approx 2.5$ pN and $K_3 \approx 0.39$ pN measured at 102.3 °C, [21] while CB6OCB possesses a uniaxial N phase in a temperature range of 110 – 157 °C. [11,22]

To change the temperature range of the $N_{tb}$ phase, nematic liquid crystal 4-pentyl-4'-cyanobiphenyl (5CB) was added to the base mixture CB7CB:CB6OCB (1:1). The nematic 5CB of 99.5 % purity was obtained from State Plant for Chemical Reagents, Kharkiv, Ukraine, and its phase transitions Cr → N at 22.5 °C and N → Iso at about 35 - 36 °C were in agreement with generally accepted values [23]. In further narration, these mixtures will be referred to as "CB-$N_{tb}$".

To induce helical twisting, two widely used ChDs were chosen, namely the (R)-2-octyl-4-[4-(hexyloxy)benzoyloxy]benzoate enantiomer known as ZLI-3786 (R-811), produced by Licrystal (Merck, Darmstadt, Germany), and cholesteryl oleyl carbonate (COC) from Aldrich, USA. [24] The enantiomer R-811 is a right-handed ChD, with phase transition Cr → Iso at 48 °C. [23] As for COC, it exhibits on cooling the Iso 36 °C CLC 22 °C Smectic-A 0 °C Cr transitions; while on heating the crystallised COC melts directly into the CLC at above 22 °C. The smectic A phase of COC is classified as monotropic. [24] In further CB-$N_{tb}$ mixtures doped with ChDs will be referred to using the acronym "CB-$N_{tb}^*$" in this study.

To obtain a planar alignment in the azimuthal plane of the CB-$N_{tb}$ (or CB-$N_{tb}^*$) mixtures, aligning films of two types were used. To study the phase transitions temperatures and textures of CB-$N_{tb}$ (or CB-$N_{tb}^*$) mixtures using a polarising optical microscope (POM), $n$-methyl-2-pyrrolidone solution of the polyimide PI2555 (HD MicroSystems, USA) in proportion 10:1 was used. The refractive index of PI2555 polyimide is $n = 1.7$. [25] For studies of optical transmission of the CB-N$_{tb}$ and CB-N$^*_{tb}$ mixtures, a polyvinyl alcohol (PVA, Aldrich) solution in water was used.

The polyimide PI2555 solution was spin coated (6000 rpm, 30 s) on glass slides. To evaporate the solvent, the polyimide PI2555 film was heated at 80 °C for 15 min and then the process of polymerisation was carried out at 180 °C for 30 min. The PI255 films were about 80 - 90 nm thick. Thereafter thin polyimide films were unidirectionally rubbed. Glass substrates, coated by polyimide PI2555, have strong azimuthal anchoring energy for number of times of unidirectional rubbings $N_{rubb} = 7 - 13$, as it was recently shown. [26]

About 100 nm thick PVA films at the glass surfaces have been formed by means of dipping technique. Pure glass substrates were dipped into prepared 1.5% water of the PVA solution followed by vertically lifting at the constant speed about 5 mm/min. To evaporate the solvent, the PVA film was heated at 100 °C for 15 min, and then the process of polymerisation was carried out at 180 °C for 90 min. The PVA films were then unidirectionally rubbed ($N_{rubb} = 5 - 8$), providing strong enough azimuthal anchoring energy for planar alignment of N and N$^*$ phases [27,28] and higher transmission in comparison with PI2555 film.

The plane-parallel symmetrical (*e.g.* sandwich-type) LC cells with the unidirectionally rubbed PI2555 (or PVA) films were assembled using two substrates with opposite rubbing directions. All LC cells were assembled using Mylar spacers, and their thickness was controlled by the interference method measuring the transmission spectrum of the empty cell by means of the Ocean Optics 4000USB spectrometer (Ocean Insight, USA). Thicknesses of LC cells were in range $d = (5 - 55) \pm 0.2$ μm.

To avoid alignment effect by flow, the LC cells were filled with CB-N$_{tb}$ (or CB-N$^*_{tb}$) mixtures by means of the capillary action at temperatures in the isotropic phase, with subsequent cooling to the N or N$^*$ state at ~ 0.5º C/min as described in. [26]

Phase transitions of mixtures and their textures were viewed through a polarising optical microscope BioLar PI (PZO, Warszawa, Poland) equipped with a digital camera Nikon D80 (Nikon, Japan). The LC cells with mixtures were placed in the thermostable heater with a temperature regulator MikRa 603 (LLD 'MikRa', Kyiv, Ukraine). The accuracy of the temperature measurement was ± 0.1°C. Temperature was measured with a platinum resistance thermometer Pt1000 (PJSC 'TERA', Chernihiv, Ukraine). The speed of the temperature change was about 0.5 - 1 °C/min.

Optical transmission spectra of CB-N$_{tb}$ (or CB-N$^*_{tb}$) mixtures were measured in sandwich-type LC cells, assembled using pair substrates with rubbed PVA films, using a Shimadzu UV-2450 spectrophotometer (Shimadzu, Japan) within spectral range 300-900 nm. The measurements were carried out within a temperature range of 17 - 77 °C for both on heating and on cooling, and the temperature values were changed in 0.2 – 0.5 °C steps and stabilized using a flowing-water thermostat (±0.1 °C).

Differential scanning calorimetry (DSC) studies were performed using a microcalorimeter Mettler DSC 1 (Mettler Toledo, Switzerland). The mixtures (~20 mg) were placed into an aluminum crucible, sealed, and thermograms were recorded in consecutive scans on heating and cooling (scanning rate 5 °C/min). The procedure was repeated 4 - 5 times. Experimental error was ± 0.1 ºC for the isotropic transition temperature and ±0.2 ºC for the transition to N$_{tb}$ phase.

**Results and discussions**

### 3.1. Differential scanning calorimetry and POM studies

The basic $N_{tb}$ mixture, consisting of CB7CB and CB6OCB in the ratio (1:1), was doped with the nematic 5CB (CB-$N_{tb}$ samples) and different ChDs (ChD-$N_{tb}$ samples) in various concentrations. The phase transition temperatures of the mixtures were determined by means of DSC and POM.

### 3.1.1. CB-$N_{tb}$ mixtures

To get an insight into the composition effects on physico-chemical properties of such mixtures, we studied changes in the $N_{tb} \rightarrow N$ and $N \rightarrow Iso$ phase transition temperatures caused by addition of various amount of the nematic 5CB in the basic CB7CB:CB6OCB mixture. The results from DSC studies are shown in Figure 1 (a). As it can be seen from Figure 1, the addition of 5CB to CB7CB:CB6OCB leads to a decrease in the temperatures of phase transitions for both $N_{tb} \rightarrow N$ and $N \rightarrow Iso$. However, the temperature range of N phase becomes wider with increasing 5CB content (Figure 1 (b)).

Since crystallization of the mixtures studied occurred with substantial and not well reproducible supercooling, we did not carry out systematic measurements below the room temperature.

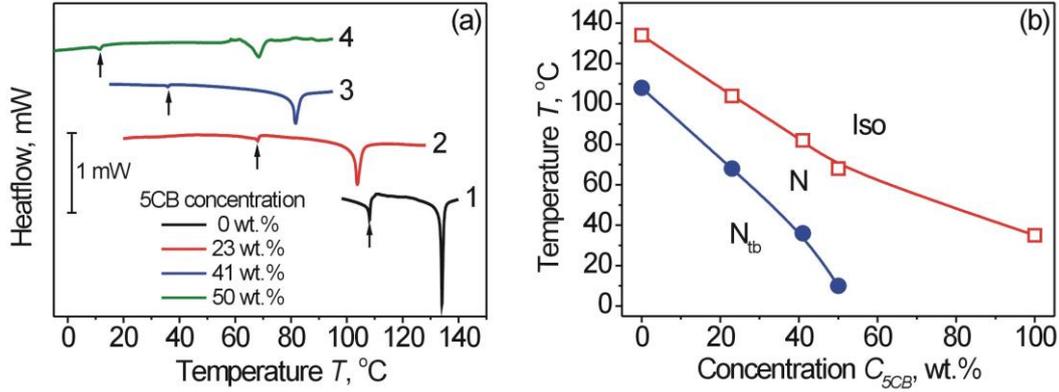

Figure 1. (a) DSC thermograms during heating of the base $N_{tb}$ mixture CB7CB:CB6OCB (1:1) upon addition of nematic 5CB ($C_{5CB}$): (1) - 0 wt.%; (2) – 23 wt.%; (3) – 41 wt.% and (4) – 50 wt.%. Arrows point to $N_{tb} \rightarrow N$ phase transitions; large endothermic peaks correspond to $N \rightarrow$ Iso phase transitions. (b) Phase transition temperatures between isotropic (Iso), nematic (N) and twist-bend nematic ($N_{tb}$) phases in CB-$N_{tb}$ mixture as function of 5CB concentration: Iso $\rightarrow$ N (open red squares) and N $\rightarrow N_{tb}$ (solid blue circles). The lines are guide to the eye.

It can be noted that the obtained concentration dependences are rather close to linearity, with a slight negative deviation of $T_i(c)$ typical for most nematic mixtures, and the barely noticeable positive deviation of the $N_{tb} \rightarrow N$ transition points. This is in agreement with similar plots for binary mixtures of $N_{tb}$-forming components [11,29]; the general appearance of DSC traces was also similar for all 5CB concentrations. Thus, it would be natural to choose the $N_{tb}$ mixture (*i.e.* CB7CB:CB6OCB in the ratio of 1:1) with addition of about 41 wt.% of the nematic 5CB as a basic CB-$N_{tb}$ system for studies of the effects of chiral dopants, since both N $\rightarrow$ Iso and N $\rightarrow N_{tb}$ transition temperatures could then be easily studied using spectrophotometers and optical microscopes with heating stages.

Photographs of the typical textures of $N_{tb}$ phase at about 23 °C in LC cells, filled by CB-$N_{tb}$ mixture with concentration of the nematic 5CB $C_{5CB}$ = 41 wt.%, placed between crossed polarizer (P) and analyzer (A) of the POM are shown in Figure 2. The typical periodic stripe texture of the $N_{tb}$ phase is clearly present and consistent with recent reports on CB-$N_{tb}$ mixtures. [22,29]

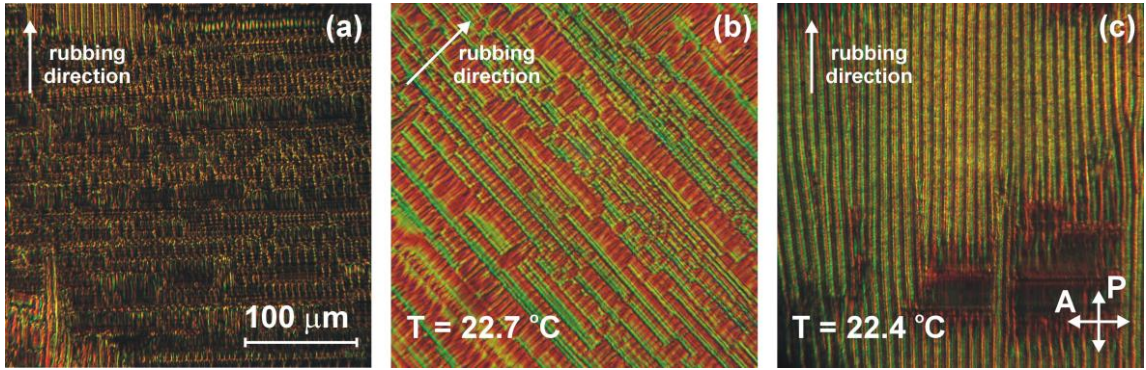

Figure 2. Photographs of the stripes texture of $N_{tb}$ phase of the CB-$N_{tb}$ mixture, containing about 41 wt.% of the nematic 5CB, filled in 13.4 μm (a), (b) and 6.4 μm (c) LC cells and placed between crossed polarizers of POM. The rubbing direction is: coinciding with polarization plane of polarizer P in (a) and (c); rotated by about 45° relative to P in (b). LC cells temperatures were 22.7 °C (a), (b) and 22.4 °C (c). LC cells were assembled with two glass substrates, covered with planar alignment layers (PI2555) with $N_{rubb}$ = 12.

In Figure 2(a), the rubbing direction of substrates (PI2555 film used for alignment of LC) was along the polarization direction of polarizer P and the LC sample in the CB-$N_{tb}$ phase is mostly dark, similarly to the N phase with the easy axis aligned along the rubbing direction. [31] When the LC sample between crossed polarizes was rotated at 45°, light is transmitted due to the birefringence (Figure 2(b)). Due to the rubbed PI2555 layers, which provide a strong and uniform planar alignment, the polygonal texture of the $N_{tb}$ phase [11] possesses stripes parallel to the rubbing direction. However, together with stripes there are also periodic undulating pseudolayers oriented perpendicular to the rubbing direction, [22] as could be seen from Figure 2(b). The main cause of defects there is the value of the azimuthal anchoring energy $W_\varphi$ of the PI2555 layers. It is well known that anchoring energy influence on the uniformity of director alignment and, as a consequence, the uniformity of undulation texture depends on the value of $W_\varphi$, as early studied for CLCs under the alternating electric field or light. [26,30] As was experimentally shown [22], that the decreasing of the LC cell thickness $d$ or changing of the nature of aligning surfaces (*i.e.,* can be regarded as a $W_\varphi$) leads to appearance of unidirectional periodic undulation textures in $N_{tb}$ phase. The uniform undulations (so-called rope-like texture [11]) of the $N_{tb}$ phase of the mixture CB-$N_{tb}$, containing 41 wt.% of the nematic 5CB, for 6.4 μm thickness LC cell is shown in Figure 2(c).

### 3.1.2. ChD-$N_{tb}$ ($N^*_{tb}$) mixtures

In order to avoid effects that could emerge in the four-component mixtures, mixtures of the basic $N_{tb}$ and various ChDs (*i.e.,* without 5CB) were studied. DSC results on these mixtures are shown in Figure 3.

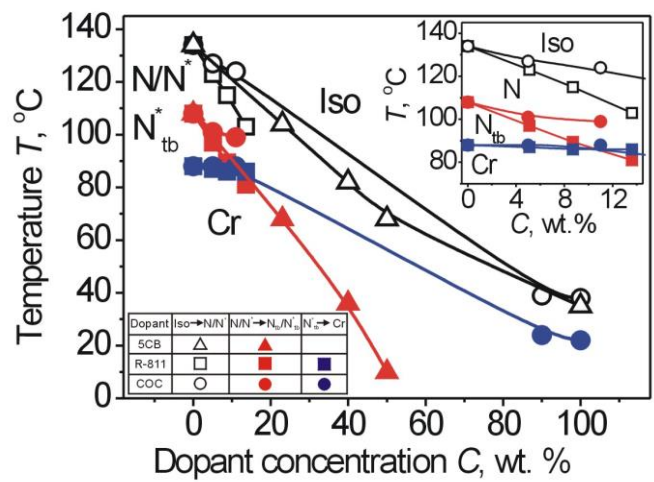

Figure 3. Phase transition temperatures between isotropic (Iso), nematic/chiral nematic (N/N*), twist-bend nematic/chiral nematic ($N_{tb}$/N*$_{tb}$) and crystal (Cr) phases in the mixture of CB7CB:CB6OCB (1:1) as function of concentration of the introduced

dopants: 5CB (open and solid triangles), R-811 (open and solid squares) and COC (open and solid circles). The insert depicts picture in the enlarged scale at low ChD concentrations. The curves are guide to the eye.

As can be seen from Figure 3, the transition temperatures decrease approximately linearly with concentration for R-811. Deviations from linearity for COC may be due just to lower solubility in the $N_{tb}$ mixture (this is in agreement with the more widened DSC peak shapes). It is interesting to note that the LC phase is supercooled for 14 wt.% R-811 of the ChD-$N_{tb}$ mixture - the $N^*_{tb}$ phase was observed only on cooling; during the heating process the crystallized mixture melts directly into the $N^*$ phase. Thus, for ChD-$N_{tb}$ mixture, containing 14 wt.% R-811, the $N^*_{tb}$ phase is monotropic.

Looking at the expanded diagram that includes all the concentration range of the ChDs, with mesogenic 5CB and COC (Figure 3) we can put down the points obtained at the opposite edge of the phase diagram, *i.e*., for 5CB and COC doped with small concentrations of the $N_{tb}$ mixture. In this case, the dependence of $T_i$ on dopant concentration $C$ is essentially close to linearity, which excludes substantial specific intermolecular interaction in the mixtures. Unfortunately, COC as ChD showed rather poor solubility in the basic $N_{tb}$ mixture CB7CB:CB6OCB in the ratio of (1:1) (though some inclination towards "banana-shaped" behavior could had been expected from its molecular structure [34]).

As can be seen from Figure 3, the phase transition temperatures in the ternary mixtures are rather high and are lowered by the introduction of dopants. In this respect, the nematic 5CB (triangle symbols) is more efficient as compared with ChDs R-811 (square symbols) or COC (circle symbols).

The textures of $N^*_{tb}$ mixture containing basic $N_{tb}$ composition (*i.e.* the ratio (1:1) of the CB7CB and CB6OCB compounds) and about 8.7 wt.% of the R-811, in a 13 μm LC cell are shown in <span style="color:red">Figure 4</span> at various temperatures, reflecting the sequence of phase transitions in this system.

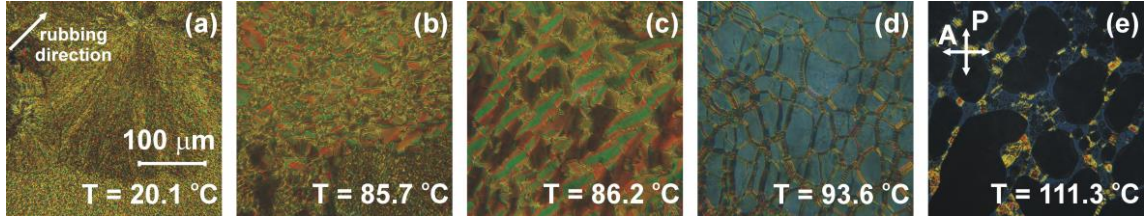

<span style="color:red">Figure 4.</span> Texture of the $N^*_{tb}$ mixture containing 8.7 wt.% of the ChD R-811 and 91.3 wt.% of the basic composition CB7CB:CB6OCB in the ratio (1:1) in 13 μm LC cell on heating: (<span style="color:red">a</span>) Cr phase at 20.1 °C; (<span style="color:red">b</span>) Cr → $N^*_{tb}$ transition at 85.7 °C; (<span style="color:red">c</span>) $N^*_{tb}$ phase with the "*platelet*" texture, at 86.2 °C; (<span style="color:red">d</span>) $N^*$ phase, Cano-Grandjean (planar) texture and oily streak defects, at 93.6 °C, and (<span style="color:red">e</span>) $N^*$ → Iso phase transition at 111.3 °C. Thickness of the LC cell was 13 μm. LC cell was placed between crossed polarizer (P) and analyzer (A) of a POM. The rubbing direction (depictured by arrow) of PI2555 layers is rotated by about 45° with respect to the polarizers.

As can be seen from <span style="color:red">Figure 4 (a)</span> and <span style="color:red">Figure 4 (b)</span>, the ChD-$N^*_{tb}$ mixture exists in the Cr phase over a wide range of temperatures (from the room temperature to ~85°C). <span style="color:red">Figure 4(c)</span> shows the $N^*_{tb}$ phase with the "*platelet*" texture at about 86 °C. When the temperature reaches 88 °C, we see the $N^*_{tb}$ → $N^*$ phase transition. Since we used the aligning layers for planar orientation of LCs, the Cano-Grandjean texture with oily streak defects characteristic of chiral nematic phase $N^*$ are observed (<span style="color:red">Figure 4(d)</span>). At 111.3 °C, the $N^*$ → Iso phase transition is observed (<span style="color:red">Figure 4(e)</span>). For this mixture Iso phase was observed at 115 °C, as can be seen from <span style="color:red">Figure 3</span> and the insert therein.

Thus, for further studies, we will concentrate on the CB-$N^*_{tb}$ mixture, including about 41 wt.% of the nematic 5CB, which will be doped with various amount of the ChD R-811.

### *3.1.3 CB-N$^*_{tb}$ mixtures*

DSC thermograms for CB-N$^*_{tb}$ mixtures consisting of CB7CB:CB6OCB (1:1) and 41 wt.% of the nematic 5CB doped by various concentration of the R-811 (*C$_{R-811}$*), are shown in <span style="color:red">Figures 5(a)</span>. The temperatures for the transitions between the isotropic (Iso), nematic (N), twist-bend/chiral twist-bend nematic (N$_{tb}$/N$^*_{tb}$) and crystal (Cr) phases in CB-N$^*_{tb}$ mixtures are shown in <span style="color:red">Figure 5b</span>

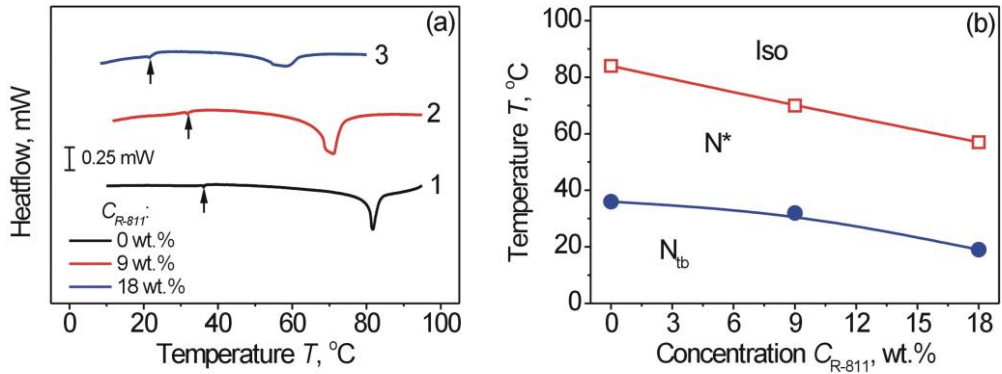

<span style="color:red">Figure 5</span>. (a) DSC thermograms during heating of CB-N$^*_{tb}$ mixtures, consisting of CB7CB:CB6OCB in the ratio of (1:1) and 41 wt.% of the nematic 5CB, doped by various concentration *C$_{R-811}$* of the ChD R-811: (1) - 0 wt.%, (2) - 9 wt.% and (3) – 18 wt.%. Arrows point to N$_{tb}$ → N (1) and N$^*_{tb}$ → N$^*$ (2,3) phase transitions, <span style="color:red">while</span> the large endothermic peaks correspond to N/N$^*$ → Iso phase transitions. (b) Phase transition temperatures between isotropic (Iso), chiral nematic (N$^*$) and twist-bend chiral nematic (N$^*_{tb}$) phases in the CB-N$^*_{tb}$ mixture as a function of *C$_{R-811}$*. The curves are guide to the eye.

As can be seen from <span style="color:red">Figure 5(a),</span> increasing the amount of ChD R-811 in the CB-N$^*_{tb}$ mixture leads to the depression of the phase transition temperatures for both the Iso → N$^*$and the N$^*$ → N$^*_{tb}$ transitions. It should be noted that the same character of changes was observed when various concentration of 5CB were added to basic N$_{tb}$ mixture, *i.e.* CB7CB:CB6OCB (1:1), as shown in <span style="color:red">Figure 1(a)</span>.

In addition, the large endothermic peak corresponding to N → Iso phase transition when CB-N*$_{tb}$ mixture does not contain ChD (black curve 1) is broadened by adding 9 wt.% and 18 wt.% of the ChD R-811, as can be seen from curves 2 and curve 3 of the Figure 5 (a), respectively.

Textures of the CB-N*$_{tb}$ mixture containing nematic 5CB with $C_{5CB}$ = 41 wt.% and ChD R-811 with $C_{R-811}$ = 18 wt.% (LC cell with PI2555 alignment layer) at various temperatures are shown in Figure 6 ((a)-(e) on heating and (f)-(j) on cooling). The 13 μm thickness LC cell was placed between crossed polarizers of POM with the rubbing direction at 45° with respect to the transmission axes of the polarizers

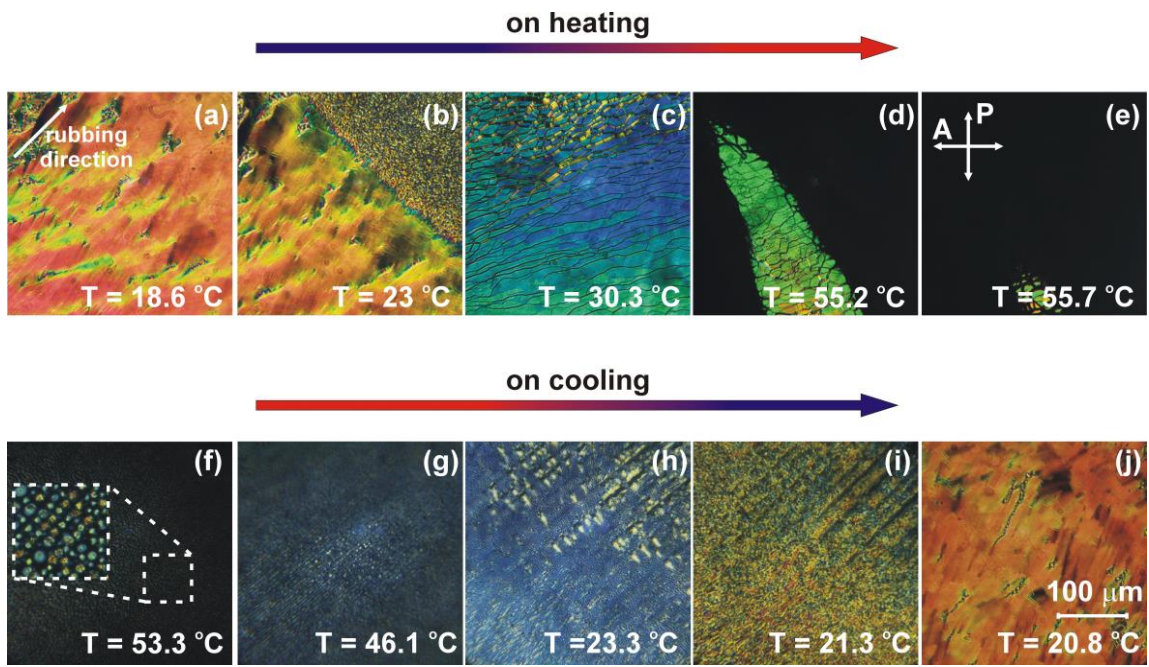

Figure 6. Textures of the CB-N*$_{tb}$ heliconical mixture containing 18 wt.% R-811 and 41 wt.% 5CB in the basic N$_{tb}$ composition CB7CB:CB6OCB (1:1), upon heating (a) – (e) and cooling (f) – (j). During the heating process: (a) the low-temperature blocky N*$_{tb}$ texture at 18.6 °C; (b) the blocky and polygonal textures at 23 °C; (c) the Cano-Grandjean (planar) texture with oily-streak defects at 30.3 °C; (d) the N* → Iso at 55.2

°C; (e) the Iso phase at 55.7 °C. During the cooling process: (f) the bubble texture of $N^*$ phase at 53.3 °C; (g) the Cano-Grandjean texture of N* phase at 46.1 °C; (h) the $N^* \rightarrow$ $N^*_{tb}$ transition and chain-like defects at 23.3 °C; (i) the polygonal texture at 21.3 °C; (j) the blocky $N^*_{tb}$ texture at 20.8 °C. The rubbing direction is rotated about 45° with respect to polarization direction of polarizer P of microscope. The thickness of LC cell is 13 μm. The bubble texture of $N^*$ phase is shown as portion of the (f) on an enlarged scale 50×50 μm².

The blocky $N^*_{tb}$ texture is observed in Figure 6 (a). During the heating process, a polygonal texture appears at about 23 °C (Figure 6 (b)). As can be seen from Figure 6 (c), on further heating, the emerging heliconical $N^*$ phase appears, showing the Cano-Grandjean (planar) texture with oily streak defects. [31]

The subsequent heating leads to the $N^* \rightarrow$ Iso phase transition (Figure 6 (d)), and the isotropic phase Iso of this heliconical mixture can be seen at 55.7 °C (Figure 6 (e)).

The cooling process is shown in Figures 6 (f)-(j). On cooling from Iso, at 53.3 °C (Figure 6 (f)) the "bubble" texture of the $N^*$ phase appears. The Cano-Grandjean texture of $N^*$ phase is observed in a wide temperature range of 52.2 – 23.3 °C (Figure 6 (g), (h)), and the short helix pitch of the $N^*$ phase allows us to observe the Bragg reflection. In a similar way as noted earlier [8, 19], at about 23.3 °C the oily-streak texture of $N^*$ phase undergoes transition to $N^*_{tb}$ (Figure 6 (h)) with appearance of a polygonal texture with chain-like defects (Figure 6 (i)). On further cooling, at about 20.8 °C the blocky $N^*_{tb}$ texture of the heliconical mixture (Figure 6 (j)) is observed.

Due to a high concentration ($C_{R-811}$ = 18 wt.%) of ChD R-811 the planar texture displays the Bragg diffraction in the visible light spectrum in a wide range of temperatures ~ 22 – 53 °C (Figure 7). The selective reflective bands, similar to those observed in conventional cholesterics, can be clearly seen. A similar set could be also obtained for a lower concentration of ChD R-811, when Bragg reflection was expected at the very edge of the visible spectral range.

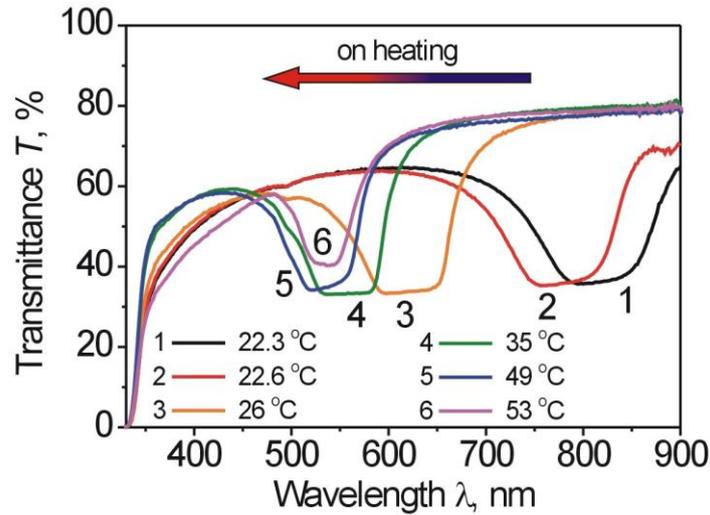

Figure 7. Transmission spectra of the CB-N$^*_{tb}$ mixture during the heating process at various temperatures in the N$^*$ phase. The CB-N$^*_{tb}$ mixture contains 18 wt.% of ChD R-811 added to CB-N$_{tb}$ mixture with 41 wt.% of nematic 5CB in basic mixture CB7CB:CB6OCB (1:1). Thickness of LC cell was about 52 μm.

At low concentration of ChD, no Bragg reflection was observed in the visible range, but the Cano-Grandjean texture was present, together with Cano disclinations characteristic for a long-pitch helix. In Figure 8, the wavelength of selective Bragg reflection maximum (λ$_{max}$) is shown as function of temperature for various concentrations of ChD R-811.

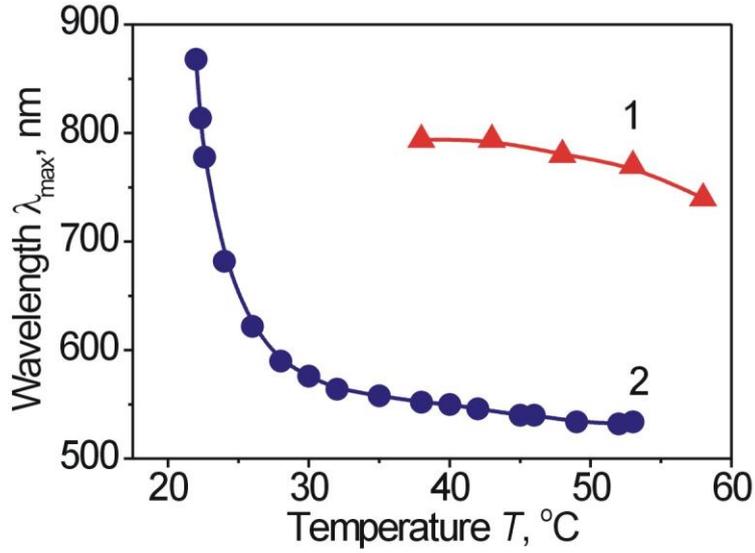

Figure 8. The wavelength of maximum selective reflection $\lambda_{max}$ as function of temperature for the CB-N*$_{tb}$ mixture comprising the nematic 5CB ($C_{5CB}$ = 41 wt.%) in the basic mixture CB7CB:CB6OCB (1:1) and various concentration of ChD R-811: (1) - $C_{R-811}$ = 9 wt.% (solid red triangles) and (2) - $C_{R-811}$ = 18 % (solid blue circles). The thickness of LC cells was about 52 μm. The curves are guide to the eye.

It is seen that in the chiral phase N* (far from N* → N$_{tb}$ phase transition) the helical pitch observed with $C_{R-811}$ = 18 wt.% corresponds to selective reflection at about 550 nm (curve 2, Figure 8), which is in agreement with the helical twisting power of R-811 for typical nematic matrices. [23] The measured values for $C_{R-811}$ = 9 wt.% of R-811, though at the very edge of the measurable region, are also consistent with that. Thus, the cholesteric helix is unwinding upon cooling, which is qualitatively similar to the well-known behaviour of cholesterics when approaching the smectic-A phase [33, 34]. In this sense, the low-temperature N*$_{tb}$ phase may cause the same behavior (i.e. suppression of helical twisting) in the pre-transitional region of the higher temperature N* phase. Thus, helical twisting above the N*$_{tb}$ → N* transition and the helical twisting in the low-temperature N*<sub>tb</sub> phase, where the helical pitch is by at least an order of magnitude lower, seem to be not correlated and appear to have quite different mechanisms of their physical origin.

For a mixture with low concentration of ChD the textures in various phases are shown in Figure 9 (CB-N$^*_{tb}$ mixture containing 20 wt.% of the nematic 5CB and 4.2 wt.% of the COC in 8.7 μm LC cell). As in Figures 4 and Figure 6, the sequence of phase transitions in the LC cell placed between crossed polarizer (P) and analyzer (A) of POM was observed.

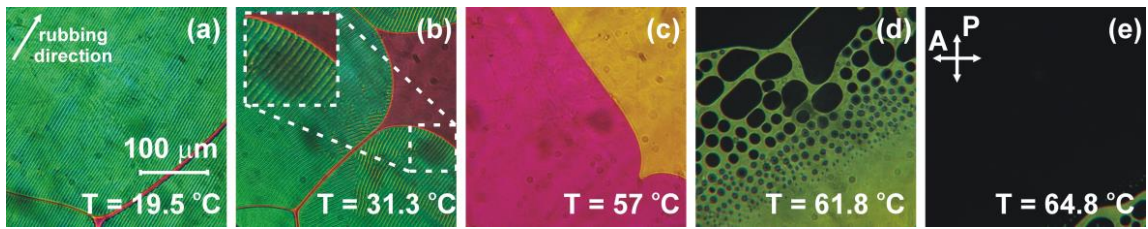

Figure 9. Texture of the CB-N$^*_{tb}$ mixture, containing 20 wt.% of the nematic 5CB and 4.2 wt.% of the COC in a 13.7 μm LC cell, during the heating process: (a) N$^*_{tb}$ phase at 19.5 °C, characterized by rope-like texture; (b) N$^*_{tb}$ → N$^*$ transition at 31.3 °C; (c) N$^*$ phase, possessing of the Cano-Grandjean (planar) texture, at 57 °C; (d) N$^*$ → Iso transition at 61.8 °C and (e) Iso phase at 64.8 °C. Thickness of the LC cell was 8.7 μm. LC cell was placed between crossed polarizer (P) and analyzer (A) of POM. The rubbing direction (depictured by arrow) of PI2555 layers is rotated about 45° relative to the polarizers. The periodical stripe texture is shown as portion of the (b) on an enlarged scale 50×50 μm$^2$.

As can be seen from Figure 9 (a), within near-room temperature range 19 – 30 °C the CB-N$^*_{tb}$ mixture forms the chiral twist-bend phase N$^*_{tb}$ with typical rope-like texture. The directions of periodical stripes (undulations) coincide with direction of the rubbing of the PI2555 layer, used as a planar aligning film. Increasing the temperature to 31.3 °C leads to the appearance of the chiral phase N$^*$ as shown in Figure 9 (b). The periodical stripes are still present, as in the inset of Figure 9 (b). The typical Cano-Granjean texture and Cano disclinations were observed in a wide temperature interval, about 32 – 60 °C (Figure 9 (c)). However, as mentioned above, the concentration of the ChD COC was only 4.2 wt.%, and consequently no Bragg diffraction in the visible spectrum was observed.

### 3.2. Optical transmission spectra studies of CB-N$^*_{tb}$ mixtures

In this section we will describe the behaviour of the optical transmission spectra measured at different temperatures in LC cells filled by CB-N$_{tb}$ or various CB-N$^*_{tb}$ mixtures with different concentrations of ChDs. The obtained values of optical transmittance will be analyzed at a certain wavelength λ that is sufficiently high to be far enough from the eventual absorption and Bragg selective reflection bands.

Among the objectives of our studies, we envisaged a comparison of different methods to characterize the N$^*_{tb}$ → N$^*$ → Iso phase transitions. In fact, the DSC method is sometimes not fully reliable for very small low-enthalpy peaks of N$^*_{tb}$ → N$^*$ transitions, especially in the case of multicomponent mixtures, and optical microscopy sometimes reflects subjective views of the observer. It seemed promising to find out how phase transitions and temperature changes of the optical transmission are interrelated. This general approach has been used in our recent studies of liquid crystals doped with carbon nanotubes and other nanoparticles [32,35,37], where pre-transitional effects were quite obvious both for N → I and N → Sm phase transitions.

Typical plots of transmission $\tau$ versus temperature $T$ obtained in multiple measurements for various mixtures comprising the basic $N_{tb}$ mixture CB7CB:CB6OCB (1:1) and different dopants are shown in Figure 10. The heating and cooling rates were ~ 0.5 C/min; the data presented were averaged over several cooling and heating cycles.

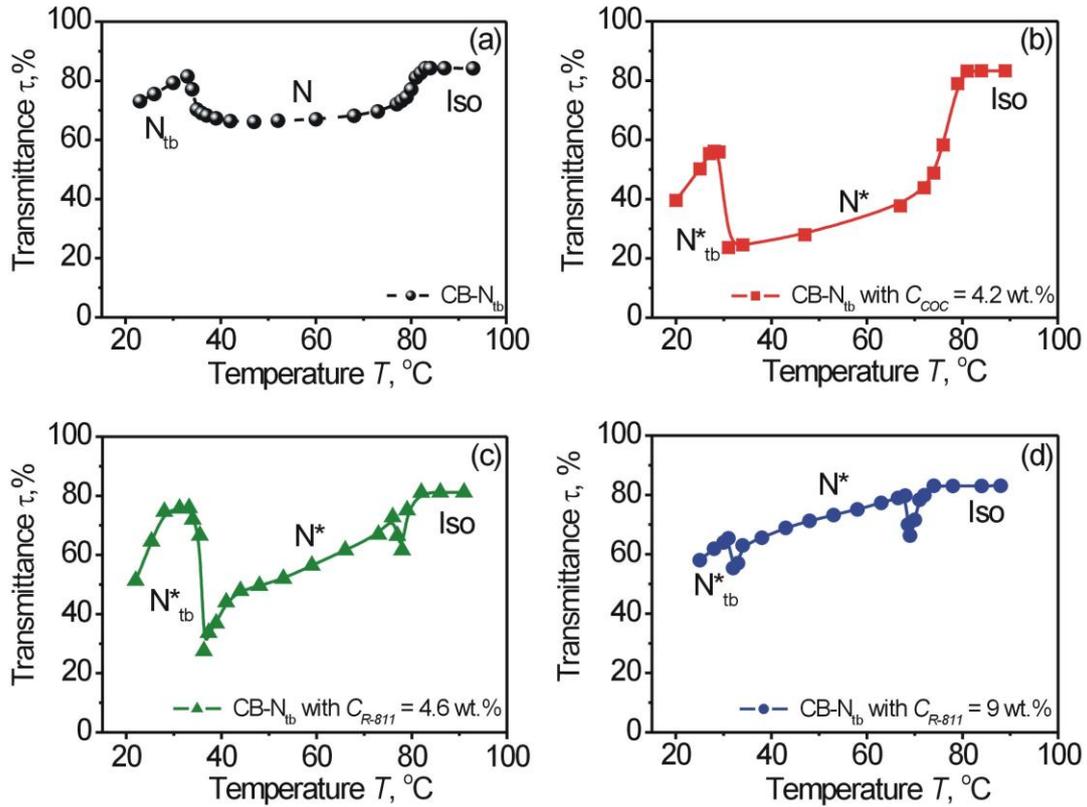

Figure 10. Transmission of LC cell at $\lambda$ = 800 nm as function of temperature $T$ for the basic mixture CB7CB:CB6OCB (1:1), containing: (a) nematic 5CB ($C_{5CB}$ = 41 wt.%) and (b-d) the mixture with 5CB doped by different ChDs; namely, (b) – 4.2 wt.% of COC; (c) - 4.6 wt.% of R-811; (d) – 9 wt.% of R-811. Thicknesses of LC cells were about $13 \pm 0.5$ µm. The curves are guide to the eye.

The $N^*_{tb} \rightarrow N^*$ and $N^* \rightarrow$ Iso phase transition temperatures estimated from the transmission curves are in good agreement with the values obtained by DSC. To exclude the effects of Bragg reflection during the measurements of the temperature dependence of transmission $\tau$ ($T$), the CB-$N^*_{tb}$ mixture containing high concentration of the R-811 ($C_{R-811}$ = 18 wt. %) was not considered.

The general trend of the optical transmission $\tau$ at $\lambda$ = 800 nm vs. temperature $T$ plots is similar to that observed in [36] for Iso $\rightarrow$ N $\rightarrow$ Sm transitions under similar experimental conditions. The transmission of LC cell became noticeably lower close to Iso $\rightarrow$ N transition. This is consistent with the presence of many small bubbles in the N or $N^*$/Iso phase during the heating or cooling process, as shown in the textures in Figure 4 (e) and Figure 6 (f). At the transition from the N/$N^*$ to $N_{tb}$/$N^*_{tb}$ phase, the optical transmission $\tau$ of LC cell increased similarly to the N $\rightarrow$ Sm transition. It is obvious that the striped textures, oily strikes defects or disclinations of $N^*$ phase, which cause substantial light scattering, disappear in the $N^*_{tb}$ phase, where a more uniform texture is observed (if we compare, *e.g.*, Figure 6 (i) with Figure 6 (j)). Thus, the $N_{tb}$/$N^*_{tb}$ phase in this respect (optical transmission) is also somewhat apparently similar to the emerging smectics upon cooling of N/$N^*$.

It was mentioned earlier that some CB-$N^*_{tb}$ mixtures exhibited a pre-transitional helix unwinding when the transition temperature to $N^*_{tb}$ was approached upon cooling (Figure 8). Recent publications [38,39] discussed the pre-transitional behavior of the elasticity constant $K_3$ in $N_{tb}$-forming systems. It is possible that the helical pitch divergence, related phenomenologically to the elasticity constant $K_2$, could be of similar nature.

**Conclusions**

We have endeavored comparative studies of several LC systems forming twist-bend phases at lower temperatures based on the well-known $N_{tb}$ mixture CB7CB:CB6OCB (1:1) in parallel experiments using several complementary methods. In particular, we used the polarizing optical microscopy (POM), differential scanning calorimetry (DSC), and measurements of temperature-dependent optical transmission outside the absorption bands.

It was shown that the basic CB7CB:CB6OCB mixture, upon addition of nematic 5CB and chiral dopants (of steroid or non-steroid nature), can change temperature ranges of the various phases ($N_{tb}/N^*_{tb}$, $N/N^*$ and Iso), decreasing the phase transition temperatures. The selective reflection in the visible range was observed in the $N^*$ phase with high concentration of ChD ($C_{R-811}$ = 9 and 18 wt.%). It was observed that in such mixtures there is a peculiar feature, namely sharp unwinding of the cholesteric helix when the temperature is lowered approaching the $N^*_{tb}$ phase, which arises analogies to the helix unwinding close to the smectic-A transition. Another aspect of this similarity was noted from the studies of changes of optical transmission during phase transition. The optical measurements are supported by optical microscopy and DSC data.

Acknowledgements

The authors thank W. Becker (Merck, Darmstadt, Germany) for his generous gift of chiral dopant R-811. Dr. A. Buchnev (Cambridge, England) who had supplied us with polymer PI2555, and Field service specialist IV V. Danylyuk (Dish LLC, USA) for his gift of some Laboratory equipment. Mesogens CB7CB and CB6OCB were generously donated by the Materials and Manufacturing Directorate of the Air Force Research Laboratory, Wright-Patterson AFB, Ohio. Our special gratitude goes to Mariacristina Rumi for useful discussions and valuable help in the manuscript preparation. This study was conducted within the Projects "Photophysics of the processes of optical radiation interaction with photorefractive, solid-state and bio-organic media" and "Effects of self-organization of functionally active nanoparticles" funded by National Academy of Sciences of Ukraine.

Figure Captions

Figure 1. (a) DSC thermograms during heating of the base $N_{tb}$ mixture CB7CB:CB6OCB (1:1) upon addition of nematic 5CB ($C_{5CB}$): (1) - 0 wt.%; (2) – 23 wt.%; (3) – 41 wt.% and (4) – 50 wt.%. Arrows point to $N_{tb} \rightarrow N$ phase transitions; large endothermic peaks correspond to $N \rightarrow Iso$ phase transitions. (b) Phase transition temperatures between isotropic (Iso), nematic (N) and twist-bend nematic ($N_{tb}$) phases in CB-$N_{tb}$ mixture as function of 5CB concentration: Iso $\rightarrow$ N (open red squares) and N $\rightarrow N_{tb}$ (solid blue circles). The lines are guide to the eye.

Figure 2. Photographs of the stripes texture of $N_{tb}$ phase of the CB-$N_{tb}$ mixture, containing about 41 wt.% of the nematic 5CB, filled in 13.4 μm (a), (b) and 6.4 μm (c) LC cells and placed between crossed polarizers of POM. The rubbing direction is: coinciding with polarization plane of polarizer P in (a) and (c); rotated by about 45° relative to P in (b). LC cells temperatures were 22.7 °C (a), (b) and 22.4 °C (c). LC cells were assembled with two glass substrates, covered with planar alignment layers (PI2555) with $N_{rubb} = 12$.

Figure 3. Phase transition temperatures between isotropic (Iso), nematic/chiral nematic (N/N$^*$), twist-bend nematic/chiral nematic ($N_{tb}$/N$^*_{tb}$) and crystal (Cr) phases in the mixture of CB7CB:CB6OCB (1:1) as function of concentration of the introduced dopants: 5CB (open and solid triangles), R-811 (open and solid squares) and COC (open and solid circles). The insert depicts picture in the enlarged scale at low ChD concentrations. The curves are guide to the eye.

**Figure 4.** Texture of the N*$_{tb}$ mixture containing 8.7 wt.% of the ChD R-811 and 91.3 wt.% of the basic composition CB7CB:CB6OCB in the ratio (1:1) in 13 µm LC cell on heating: (a) Cr phase at 20.1 °C; (b) Cr → N*$_{tb}$ transition at 85.7 °C; (c) N*$_{tb}$ phase with the "*platelet*" texture, at 86.2 °C; (d) N* phase, Cano-Grandjean (planar) texture and oily streak defects, at 93.6 °C, and (e) N* → Iso phase transition at 111.3 °C. Thickness of the LC cell was 13 µm. LC cell was placed between crossed polarizer (P) and analyzer (A) of a POM. The rubbing direction (depictured by arrow) of PI2555 layers is rotated by about 45° with respect to the polarizers.

**Figure 5.** (a) DSC thermograms during heating of CB-N*$_{tb}$ mixtures, consisting of CB7CB:CB6OCB in the ratio of (1:1) and 41 wt.% of the nematic 5CB, doped by various concentration $C_{R-811}$ of the ChD R-811: (1) - 0 wt.%, (2) - 9 wt.% and (3) – 18 wt.%. Arrows point to N$_{tb}$ → N (1) and N*$_{tb}$ → N* (2,3) phase transitions, while the large endothermic peaks correspond to N/N* → Iso phase transitions. (b) Phase transition temperatures between isotropic (Iso), chiral nematic (N*) and twist-bend chiral nematic (N*$_{tb}$) phases in the CB-N*$_{tb}$ mixture as a function of $C_{R-811}$. The curves are guide to the eye.

**Figure 6.** Textures of the CB-N*$_{tb}$ heliconical mixture containing 18 wt.% R-811 and 41 wt.% 5CB in the basic N$_{tb}$ composition CB7CB:CB6OCB (1:1), upon heating (a) – (e) and cooling (f) – (j). During the heating process: (a) the low-temperature blocky N*$_{tb}$ texture at 18.6 °C; (b) the blocky and polygonal textures at 23 °C; (c) the Cano-Grandjean (planar) texture with oily-streak defects at 30.3 °C; (d) the N* → Iso at 55.2 °C; (e) the Iso phase at 55.7 °C. During the cooling process: (f) the bubble texture of N* phase at 53.3 °C; (g) the Cano-Grandjean texture of N* phase at 46.1 °C; (h) the N* →

N$^*_{tb}$ transition and chain-like defects at 23.3 °C; (i) the polygonal texture at 21.3 °C; (j) the blocky N$^*_{tb}$ texture at 20.8 °C. The rubbing direction is rotated about 45° with respect to polarization direction of polarizer P of microscope. The thickness of LC cell is 13 μm. The bubble texture of N$^*$ phase is shown as portion of the (f) on an enlarged scale 50×50 μm$^2$.

Figure 7. Transmission spectra of the CB-N$^*_{tb}$ mixture during the heating process at various temperatures in the N$^*$ phase. The CB-N$^*_{tb}$ mixture contains 18 wt.% of ChD R-811 added to CB-N$_{tb}$ mixture with 41 wt.% of nematic 5CB in basic mixture CB7CB:CB6OCB (1:1). Thickness of LC cell was about 52 μm.

Figure 8. The wavelength of maximum selective reflection $\lambda_{max}$ as function of temperature for the CB-N$^*_{tb}$ mixture comprising the nematic 5CB ($C_{5CB}$ = 41 wt.%) in the basic mixture CB7CB:CB6OCB (1:1) and various concentration of ChD R-811: (1) - $C_{R-811}$ = 9 wt.% (solid red triangles) and (2) - $C_{R-811}$ = 18 % (solid blue circles). The thickness of LC cells was about 52 μm. The curves are guide to the eye.

Figure 9. Texture of the CB-N$^*_{tb}$ mixture, containing 20 wt.% of the nematic 5CB and 4.2 wt.% of the COC in a 13.7 μm LC cell, during the heating process: (a) N$^*_{tb}$ phase at 19.5 °C, characterized by rope-like texture; (b) N$^*_{tb}$ → N$^*$ transition at 31.3 °C; (c) N$^*$ phase, possessing of the Cano-Grandjean (planar) texture, at 57 °C; (d) N$^*$ → Iso transition at 61.8 °C and (e) Iso phase at 64.8 °C. Thickness of the LC cell was 8.7 μm. LC cell was placed between crossed polarizer (P) and analyzer (A) of POM. The rubbing direction (depictured by arrow) of PI2555 layers is rotated about 45° relative to the polarizers. The periodical stripe texture is shown as portion of the (b) on an enlarged scale 50×50 μm².

Figure 10. Transmission of LC cell at λ = 800 nm as function of temperature $T$ for the basic mixture CB7CB:CB6OCB (1:1), containing: (a) nematic 5CB ($C_{5CB}$ = 41 wt.%) and (b-d) the mixture with 5CB doped by different ChDs; namely, (b) – 4.2 wt.% of COC; (c) - 4.6 wt.% of R-811; (d) – 9 wt.% of R-811. Thicknesses of LC cells were about 13 ± 0.5 μm. The curves are guide to the eye.